\documentstyle[prd,aps,preprint,tighten]{revtex}

\begin{document}
\draft

\title{$Z_3$ Flavor Symmetry and Possible Implications}
\author{ {M.J.Luo}\thanks{E-mail: mjluo@mail.ustc.edu.cn}\\
{\sl Department of Modern Physics,\\
University of Science and Technology of China,\\
Hefei, Anhui 230026, China}\\
} \maketitle

\begin{abstract}
We show in this paper that the $Z_3$ flavor symmetry, which can
successfully produce the tri-bimaximal mixing and flavor pattern of
neutrino sector, has a possible explanation in the framework of
gauge symmetry by constructing a wavefunction of flavor state
particles with the help of the Wilson loop. In this implementation
of $Z_3$ flavor symmetry, we suggest that the flavor charge in weak
interaction can be interpreted as a topological charge. Its possible
implications and generalizations to the quark sector are also
discussed.
\end{abstract}

\section{Introduction}
The Standard Model describes almost all laboratory data with 28 free
parameters, most of them arise from the flavor and mass parameters
in the Yukawa coupling $y_{ij}$ between fermions and Higgs boson
$H$, the Lagrangian is
\begin{eqnarray}
{\cal L}_{Yukawa}= y^u_{ij} \bar{u}_i q_j H + y^d_{ij} \bar{d}_i q_j
H^* + y^e_{ij} \bar{e}_i l_j H^*.
\end{eqnarray}
where $q,l,({\bar u},{\bar d},{\bar e})$ are left (right) handed
quarks and leptons. Although 17 in 22 of the flavor parameters are
measured\cite{PDG}, to understand these free parameters is a great
challenge. The dominant approach is constructing flavor symmetry to
reduce the number of free parameters, e.g \cite{F-N}.

Beyond the Standard Model, neutrino oscillation
experiments\cite{neutrino}, give us strong evidences that the
neutrino also have non-zero masses and non-trivial mixing between
mass eigenstates and flavor states
\begin{eqnarray}
\nu_{\alpha} = \sum_{a=1}^3 U_{\alpha a} {\nu_a},
\end{eqnarray}
in which $a$ denotes the mass eigenstate and $\alpha$ the flavor
state, $U_{\alpha a}$ is the MNS matrix \cite{MNS} that has the form of nearly
tri-bimaximal\cite{TB}.

The request of understanding the tri-bimaximal mixing demands a
theory of flavor, and the neutrino masses and mixing matrix have
inspired many model buildings, e.g. non-Abelian discrete flavor
symmetries\cite{A4}, GUT$\times$discrete group models\cite{GUT},
shift symmetries\cite{shift} etc., in which a heuristic model is the
Abelian $Z_3$ flavor symmetry, e.g.\cite{z3}.

The observational facts that the mixing between mass eigenstates and
flavor states, as well as the universality of the flavor states in
weak interaction imply that the elementary excitations of weak
process are non-Fock \cite{nonfock}, and have non-trivial
structures. As it is known that the Fock quantization is closely
related to the particle interpretation of non-interacting QFT in
which momentum, energy (mass) and spin of particle are good quantum
numbers. However, the flavor state does not carry definite mass and
spin as its quantum numbers, but definite flavor charge which is
instead seem as good quantum number for weak interaction. Actually,
the flavor states are the eigenstates of the weak interaction and
the Haag's theorem \cite{haag} states that Fock state does not exist
for interacting QFTs.

Such non-Fock degrees of freedom (DoF) may be crucial for
understanding the mixing phenomenon in weak processes. We suggest
that the $Z_3$ flavor symmetry model for neutrino mixing is very
heuristic, since the Wilson loop operator has a natural
implementation of $Z_3$ symmetry and can be used as a guidance to
construct the non-Fock elementary excitations. In this paper, we
will give a possible explanation of $Z_3$ symmetry in the framework
of gauge symmetry with the help of Wilson loop by introducing it to
each particle wavefunction. Then according to the non-Fock
wavefunction, the eigenvalues are not related to the mass and spin,
but rather a winding number of Wilson loop, so we could give the
flavor charge a possible interpretation.

The paper is organized as follows. We review a simple $Z_3$ flavor
symmetry for neutrino in section II, our implementation of the $Z_3$
flavor symmetry is in section III by using the Wilson loop, we give
a topological quantum number interpretation to the flavor charge.
The generalization to the quark sector, which will reproduce the
Froggatt-Nielsen's scenario for the Yukawa couplings, is discussed
in section IV.

\section{$Z_3$ flavor symmetry and neutrino mixing}
In this section, we review one of a simple and heuristic model of neutrino mixing
based on the $Z_3$ flavor symmetry. Consider the $Z_3$ elements
($\omega, \omega^2, 1$), where $\omega=e^{\frac{2 \pi i}{3}}$. We
assume neutrinos are Majorana particles, and a general Lagrangian of Majorana mass
term is
\begin{eqnarray}\label{neutrinoL}
{\cal L} = y \bar{\nu^c} \Phi \nu,
\end{eqnarray}
where $y$ is a coupling constant, $c$ stands for the charge
conjugation $\nu^c=C {\bar\nu^T}$. Since $Z_3$ is a one-dimensional
Abelian group, so the transformation is purely multiplying an
imaginary phase, $\omega^i \in Z_3$. The Lagrangian is invariant
under the $Z_3$ transformation
\begin{eqnarray}\label{z3trans}
\nu_i \rightarrow \omega^i \nu_i, \nonumber\\
\bar{\nu_i^c} \rightarrow \bar{\nu_i^c} \omega^i, \nonumber\\
\phi_i \rightarrow \omega^i \phi_i,
\end{eqnarray}
where the index $i$ from 1 to 3, three Higgs fields are introduced in the
model. After the transformation, the coupling takes the form
\begin{eqnarray}\label{yukawa}
y (\omega^i \omega^j \omega^k) \bar{\nu^c_i} \phi_j \nu_k,
\end{eqnarray}
where $\omega^i \omega^j \omega^k$ in the parentheses should be an
invariant of $Z_3$, so it leads to $i+j+k=0 \,\, mod \,\, 3$, which
constrains the relations between $i,j,k$. Expanding it into matrix
in flavor basis, the texture of coupling then has the form
\begin{eqnarray}\label{massmatrix}
y \left( \matrix{\bar{\nu^c_1} & \bar{\nu^c_2} & \bar{\nu^c_3}} \right)
\left(
\matrix{\phi_1 & \phi_3 & \phi_2 \cr
\phi_3 & \phi_2 & \phi_1 \cr
\phi_2 & \phi_1 & \phi_3}
\right)
\left(
\matrix{\nu_1 \cr \nu_2 \cr \nu_3} \right).
\end{eqnarray}
It is easy to verify that the mass matrix can be almost diagonalized
by the tri-bimaximal mixing matrix
\begin{eqnarray}
U_{tb} = \left ( \matrix{ \frac{2}{\sqrt{6}} &
 \frac{1}{\sqrt{3}} & 0 \cr
-\frac{1}{\sqrt{6}} &  \frac{1}{\sqrt{3}} &
 \frac{1}{\sqrt{2}} \cr
-\frac{1}{\sqrt{6}} &  \frac{1}{\sqrt{3}} &
 -\frac{1}{\sqrt{2}} \cr } \right ).
\end{eqnarray}
We have
\begin{eqnarray}
U^\dagger_{tb} \Phi U_{tb} = \left(
\matrix{\phi_1-\frac{\phi_2}{2}-\frac{\phi_3}{2} & 0 &
\frac{\sqrt{3}}{2}(\phi_3-\phi_2) \cr 0 & \phi_1+\phi_2+\phi_3 & 0
\cr \frac{\sqrt{3}}{2}(\phi_3-\phi_2) & 0 & -
\phi_1+\frac{\phi_2}{2}+\frac{\phi_3}{2}} \right),
\end{eqnarray}
which is diagonalized when $\phi_2 = \phi_3$. The neutrino mass
matrix is obtained by developing VEV for $\phi_i$. Because of the
non-zero (1,3) and (3,1) elements, the matrix needs further
diagonalization which will give a deviation from the tri-bimaximal
matrix and leads to a non-vanishing $\theta_{13}$. The deviation or,
equivalently, the non-vanishing $\theta_{13}$ will depend on the
magnitude of the non-zero (1,3) and (3,1) elements, i.e. the
difference of the VEV of $\phi_2$ and $\phi_3$. Here, the VEVs of
Higgs fields $\langle \phi_i \rangle$ does not necessary to be the
electroweak scale for the masses of neutrino are small, in some
models they could be a tiny scale, e.g. in Higgs triplet model
$\langle \phi_i \rangle \sim 10^{-3}eV$.

In the following sections, we will implement the transformation
properties Eq.(\ref{z3trans}) of the discrete $Z_3$ symmetry in the
framework of continuous gauge symmetry and construct similar $Z_3$
invariants as Eq.(\ref{yukawa}) to be the Yukawa couplings.

\section{Implementation of $Z_3$ symmetry and Possible Implications}
Now in general we discuss the discrete $Z_N$ symmetry from the
continuous gauge $SU(N)$ symmetry. The possible implications of
$Z_N$ flavor symmetry will be a good guidance for us to find the
non-Fock wavefunction of the flavor state.

As it is well known that the gauge transformation on a particle, whose operator is
near the identity element of gauge group $G=SU(N)$ (we call it small
gauge transformation), it transforms as a fundamental
representation of $G$,
\begin{eqnarray}
\Psi \rightarrow e^{i \theta^a t^a} \Psi \simeq (1+i \theta^a t^a) \Psi,
\end{eqnarray}
where $t^a$ is the generator of G, $\theta^a$ is the parameter of
transformation and $\Psi$ is the wavefunction of the particle. A
gauge potential $A=e A^a_{\mu} t^a dx^{\mu}$ induces a finite phase to the
particle when it passes along a path $l(x,x^\prime)$ from $x$ to
$x^\prime$,
\begin{eqnarray}
\Psi(x^\prime) = e^{i \int_{l(x,x^\prime)} A} \Psi(x).
\end{eqnarray}
The gauge potential $A$ transforms as
\begin{eqnarray}\label{Atrans}
A \rightarrow A^g = g^{-1} A g + \frac{1}{i}g^{-1} d g = A + \frac{1}{i} g^{-1} D g,
\end{eqnarray}
where $g \in G$ defines a mapping from the base manifold to the
gauge group, $g(x) = e^{i \theta^a(x) t^a}$, i.e. $g(x): M
\rightarrow G$, and $D$ is the covariant derivative $D=d+iA$. If the
mapping $g$ is non-trivial, then an extra phase appears which
related to the gauge transformation whose operator does not contain
the identity element, in other words, the gauge transformation has
non-zero "winding number" and we call it large gauge transformation.
We have
\begin{eqnarray}\label{decomp}
\Psi(x^\prime) = e^{i \int_{l(x,x^\prime)} A} e^{\int_{l(x,x^\prime)} g^{-1} D g} \Psi(x),
\end{eqnarray}
it gives a decomposition of a small gauge perturbation and large
gauge transformation. The former responds to the local phase for a
Fock-like particle, while the latter phase needs an extra quantum
number to describe which has no effect on the local phenomenon.

When we consider that the path is closed to be a loop $\gamma_x$ at
a base point $x$, the first phase factor tends to vanish as the
closed loop shrinks to the point $x$, while the second one remains
for the obstacle in the non-simply connected space,
\begin{eqnarray}
\Psi(x')|_{x' \rightarrow x} \rightarrow e^{\oint_{\gamma_x} g^{-1} D g} \Psi(x).
\end{eqnarray}
Obviously, the trace of phase factor, which does not rely on the
choice of the basis of the gauge group, is an invariant function
under gauge transformation and independent with the spacetime
metric, so it is expected to be an observable and be part of the
wavefunction. The general form of the physical phase is written as
the so-called Wilson Loop \cite{wilsonloop}
\begin{eqnarray}
W_\gamma[A^g]=tr ( P e^{i \oint_\gamma A^g}),
\end{eqnarray}
where $P$ denotes the path ordering along $\gamma$.

When the mapping $g$ is non-trivial for some
topological obstacle exist, e.g. all scalar fields form
$Z_N$-vortices, the relevant gauge group $G=SU(N)/Z_N$ is not simply
connected,
\begin{eqnarray}
\pi_1 (SU(N)/Z_N) = Z_N,
\end{eqnarray}
for the mapping $g(x): S^1 \rightarrow SU(N)/Z_N$, then $g$ becomes
multivalued, consider the closed loop $\gamma$ parametrized by an
angle $\theta$ with $0 \leq \theta \leq 2 \pi$, and $\gamma(2
\pi)=\gamma(0)$, we have $g(2 \pi) = e^{2 \pi n i/N} g(0)$ with $0
\leq n < N$. We say that the field has a winding number $n$ in such
a configuration.

As a consequence, there does not exist a global section to be the
wavefunction in the case that the bundle is topological non-trivial,
and each section differs by an extra phase representing a large
gauge transformation. So according to the decomposition
Eq.(\ref{decomp}), the wavefunction of particle can be represented
as a trivial wavefunction $\Psi_0(x)$ multiplying an extra phase
exhibited by the non-contractible Wilson loop at the base point $x$
which characterized its topological class. In general each
topological class of wavefunction can be constructed as
\begin{eqnarray}\label{ee}
\Psi(x) = W_{\gamma_x}[A] \Psi_0(x).
\end{eqnarray}
Under the gauge transformation Eq.(\ref{Atrans}), we note that
\begin{eqnarray}
W_\gamma[A^g]=tr e^{\oint_{\gamma} g^{-1} d g} W_\gamma [A] = e^{2 \pi n(\gamma) i/N} W_\gamma[A],
\end{eqnarray}
where $n(\gamma)$ is the number of times the loop $\gamma$ winding
around the obstacle, which is topological stable against small
perturbations,
\begin{eqnarray}\label{stable}
n_\gamma[A]=n_\gamma[A+\delta A].
\end{eqnarray}

Even if the $SU(N)$ symmetry is broken completely, the phase factor
still values on the residual center of $SU(N)$, the $Z_N$, so we
assume in this paper, only the $Z_N$ DoF are relevant to the
wavefunction, which is transformed as the representation of $Z_N$
group,
\begin{eqnarray}\label{wilsontrans}
\Psi(x) \rightarrow \Psi'(x) = W_\gamma[A^g] \Psi_0(x) = e^{2 \pi n i/N} W_\gamma[A] \Psi_0(x)= e^{2 \pi n i/N} \Psi(x),
\end{eqnarray}
where $e^{2 \pi n i/N}$ is the element of $Z_N$.

It is direct to check that the wavefunction Eq.(\ref{ee}) is the
eigen-function of the Dirac operator ${D\kern-1.4ex
/}(A^g)={\partial\kern-1.4ex /}+ i{A\kern-1.4ex /}^g$. As it is well
known that the gauge transformation of ${D\kern-1.4ex /}(A)$,
\begin{eqnarray}
g^{-1}(x) {D\kern-1.4ex /}(A) g(x) = {D\kern-1.4ex /}(A^g),
\end{eqnarray}
consequently, every eigen-function $\Psi$ of ${D\kern-1.4ex /}(A)$,
\begin{eqnarray}
{D\kern-1.4ex /}(A) \Psi = \lambda \Psi,
\end{eqnarray}
has an associated gauge transformed eigen-function $g \Psi$,
\begin{eqnarray}
{D\kern-1.4ex /}(A^g) g \Psi = \lambda g \Psi,
\end{eqnarray}
so is $W[A^g] \Psi$,
\begin{eqnarray}
{D\kern-1.4ex /}(A^g) W[A^g] \Psi = \lambda W[A^g] \Psi,
\end{eqnarray}
since $W[A^g]$ valued on the center of group element $g$, without
losing generality, taking the pure gauge $A^g=g^{-1} d g$, we have
\begin{eqnarray}
W_\gamma[A^g]=tr(e^{\oint_{\gamma} g^{-1} d g}) = tr(g(2 \pi) g^{\dagger}(0))
\end{eqnarray}

In summery, if non-contractible loops exist (i.e. when the $\pi_1(G) \neq 0$ due to topological obstacle), the
wavefunction of elementary excitation can be constructed as a
topological trivial wavefunction multiplied by a
non-trivial Wilson loop that exhibits its extra topological quantum
number. So in this case, we suggest possible connections between such extra
topological DoF and the flavor DoF:

1) The wavefunction Eq.(\ref{ee}) is the eigen-function of the Dirac
operator, so it is the eigenstate of interaction similar with the
flavor state, but rather the energy eigenstate. The eigenvalue $e^{2
\pi Q i/N}=\omega^Q$ classifies the equivalence class of the
wavefunction, where the winding number measures topological charges
$Q$ that similar with the flavor indices $i$ of $\omega^i$ in
Eq.(\ref{z3trans}).

2) The flavors have similar formal behavior with the topological DoF
under $Z_N$ symmetry Eq.(\ref{wilsontrans}), the $Z_N$
transformation is interpreted as a large gauge transformation and
has no effects on local process. The local gauge quantum numbers
assigned by gauge group for different flavors are the same.

3) The flavor charge seems stable against local gauge interactions similar with
the behavior of Eq.(\ref{stable}), local flavor changing processes
such as Lepton Flavor Violation (LFV) and Flavor Changing
Neutral Current (FCNC) are highly suppressed in our observation by
far.

4) Note that if we reverse the orientation of the loop $\gamma$ and
exchange the representation of $G$ with its complex conjugate, the
definition of Wilson loop is unchanged, it is equivalent to take the
charge conjugation of the wavefunction.

These are important formal properties of the flavor DoF and the
topological DoF of the non-Fock wavefunction Eq.(\ref{ee}). The
wavefunction will have non-trivial consequences when we dealing with
the VEVs of the Wilson loops in which the connection $A$ is seen as
dynamic field operators. In the following discussing, we will base
on the non-trivial wavefunction Eq.(\ref{ee}) and take $N=3$.

\section{The Quark Sector}

In this section, we will generalize the previous discussion to
quantum version by seeing the wavefunction Eq.(\ref{ee}) as a field
operators. In the standard treatment of effective QFT, the high
energy DoF will be integrated out and contribute to the effective
low energy DoF, we will find that the VEVs of the Wilson loops in
particle's wavefunction will play a crucial role in their effective
couplings which give a natural realization of strong hierarchy of
quark masses.

First we briefly reconsider the neutrino case. Assume that the
wavefunction of neutrino is almost identified with the trivial
wavefunction $\Psi \simeq \Psi_0$, that is $\langle W[A] \rangle
\simeq 1$, the VEV of the transformed Wilson loop in the neutrino
wavefunction is then purely an imaginary phase valued on $Z_3$,
\begin{eqnarray}
\langle W[A] \rangle \rightarrow \langle W[A^g] \rangle=\omega^n \langle W[A] \rangle \simeq \omega^n \in Z_3,
\end{eqnarray}
in other words, the wavefunction of neutrinos are the eigenstates of
the Wilson loop operators with eigenvalues $\omega^n$. Hence the
result is the same as the section II. The Lagrangian takes the
similar form of Eq.(\ref{massmatrix}), only the Higgs fields develop
non-vanished VEVs with the scale of, e.g. in Higgs triplet model,
$\langle \phi_i \rangle \sim 10^{-3}eV$.

However, when we consider the quark sector, the quantum expectation
values of Wilson loops will not be trivial, if the local DoF of
quarks are confined. Writing the Yukawa coupling terms in ordinary
form, where the gauge potential in the Wilson loop operators should
be integrated out and the VEVs of such loop operators then make
contributions to the effective Yukawa couplings, they formally
become the correlation function of these Wilson loop operators,
which are observables that gauge invariant under $SU(3)/Z_3$. They
require their VEVs and the Lagrangian of quark mass terms take the
form,
\begin{eqnarray}
\langle W_{q_{aL}}^\dagger W_{H_c} H_c W_{q_{bR}}
\rangle \bar{q}_{aL} q_{bR},
\end{eqnarray}
where the subscript $a,b,c$ of the Wilson loop are integers
representing the corresponding winding number of the loop, or
equivalently in this paper, the flavor species, we write it as
\begin{eqnarray}
W_{n_a} = tr( P \exp{i \oint_{\gamma(n_a)} A} ),
\end{eqnarray}
where $n_a$ is the winding number. Taking the number $\langle H_c
\rangle$ out of the bracket, we have the expectation value of the
correlation function of these three Wilson loops, which can be
calculated by the standard Feynman path-integral, formally it can be
written as
\begin{eqnarray}\label{www}
\langle W_{a}^\dagger W_{c} W_{b} \rangle = Z^{-1} \int D A
W_{a}^\dagger W_{c} W_{b} e^{-S[A]}.
\end{eqnarray}
The symbol $DA$ represents Feynman's integral over all gauge orbits,
that is, all equivalence classes of connections modulo gauge
transformations, and $S[A]$ is the action of gauge theory in 4
dimension with gauge group $G=SU(3)/Z_3$. The approximate behaviors
of Eq.(\ref{www}) can be given as follows. We assume that the loop
operators are almost independent, the expectation value can be
decomposed as
\begin{eqnarray}
\langle W_{a}^\dagger W_{c} W_{b} \rangle \simeq \langle
W_{a}^\dagger \rangle \langle W_{c} \rangle \langle W_{b} \rangle e^{i \delta(\gamma_{a},\gamma_{c},\gamma_{b})},
\end{eqnarray}
in which $e^{i \delta(\gamma_{a},\gamma_{c},\gamma_{b})}$ is an observable
angle depending on three Wilson loops. A crucial properties of
$\langle W_\gamma \rangle$ is that if the DoF of quarks are confined it
has area law \cite{thooft},
\begin{eqnarray}
\langle W_{n_\gamma} \rangle = tr \exp \left( - \oint_{\gamma}
dx^{\mu} \oint_{\gamma} dy^{\nu} \langle A_{\nu}(y) A_{\mu}(x)
\rangle \right) \simeq e^{- \sigma {\cal A}_\gamma},
\end{eqnarray}
where $\langle A_{\nu}(y) A_{\mu}(x) \rangle$ is the propagator,
${\cal A}_\gamma$ is the area of surface whose boundary is the loop,
it is approximate that the area, which the flux goes through,
increases with the winding number, i.e. ${\cal A}_\gamma \simeq
n_\gamma {\cal A}$, and $\sigma$ is a fixed constant.


It gives quark a strong mass hierarchy. Neglecting the complex phase
we have the Yukawa couplings
\begin{eqnarray}
y_{ab} \simeq g_{H} e^{-\sigma (n_a {\cal A}_{q_{L}}+ n_b {\cal A}_{q_{R}})},
\end{eqnarray}
where only one Higgs field is involved and the $\langle W_H \rangle$
contributes to the coupling $g_H$, the VEV of Higgs field is of the
electroweak scale $\langle H \rangle \simeq 246 GeV$.

Comparing it with the form of the Yukawa couplings resulting from
approximate Froggatt-Nielson\cite{F-N} Abelian $U(1)$ Flavor Symmetries, in
which the general scheme is a small symmetry breaking factor for
each quark field, $\epsilon_{q,\bar{u},\bar{d}} \simeq \frac{\langle \theta_{FN} \rangle}{\Lambda}$ that leads to the
Yukawa coupling elements
\begin{eqnarray}
y^u_{ik}=g \epsilon_{\bar{u}}^i \epsilon_q^k, \,\,
y^d_{jk}=g^\prime \epsilon_{\bar{d}}^j \epsilon_q^k,
\end{eqnarray}
in which $\langle \theta_{FN} \rangle$ is the VEV of an introduced
field, $\Lambda$ an energy scale and $i,j,k$ are the integer flavor
indices. Therefore, in our scenario it is natural to have the
identification
\begin{eqnarray}
\epsilon_{\bar{u}}^{i} = e^{-\sigma n_a {\cal A}_{u_{L}}}, \,\,
\epsilon_{\bar{d}}^{j} = e^{-\sigma n_b {\cal A}_{d_{L}}}, \,\,
\epsilon_{q}^{k} = e^{-\sigma n_c {\cal A}_{q_{R}}},
\end{eqnarray}
where the integer flavor indices $i,j,k$ identify with the winding
number $n_a,n_b,n_c$. So it is a possible scenario to fit the flavor
pattern of quark sector well.

\section{Conclusion}
In this paper, we discussed a simple and heuristic $Z_3$ flavor
symmetry model for neutrino masses, and gave it a possible
realization by constructing a non-Fock wavefunction involving the
Wilson loop, the Eq.(\ref{ee}). In this scenario, we show that the
flavor charge can be interpreted as topological charge. The flavor
DoF has similar behavior with the topological DoF under $Z_3$
symmetry, and it is stable against local small gauge perturbation
which is consistent with the fact that flavor changing processes are
suppressed in our observations by far.

A possible generalization to the quark sector is also discussed, in
which we gave a possible scheme to compute the Yukawa couplings for
quarks by calculating the correlation function of the Wilson loops
in their wavefunctions. This scheme leads to strong hierarchy for
the quark Yukawa couplings and reproduce the similar textures from
the scenario of Froggatt-Nielson's Abelian Flavor Symmetries.


\end{document}